\documentclass[pra,twocolumn,showpacs,superscriptaddress]{revtex4}

\usepackage{amsmath}
\usepackage{amssymb}

\begin{document}
\bibliographystyle{apsrev}

\title{An entanglement monotone derived from Grover's algorithm}
\date{\today}

\author{Ofer Biham}
\affiliation{Centre for Quantum Computer Technology and
Department of Physics, University of Queensland, Brisbane,
Queensland 4072, Australia} 
\affiliation{Racah Institute of
Physics, The Hebrew University, Jerusalem 91904, Israel}
\author{Michael A. Nielsen}
\affiliation{Centre for Quantum Computer Technology and
Department of Physics, University of Queensland, Brisbane,
Queensland 4072, Australia} 
\affiliation{Institute for Theoretical
Physics, Kohn Hall, \\
University of California Santa Barbara, Santa
Barbara, CA 93106, USA}
\author{Tobias J. Osborne}
\affiliation{Centre for Quantum Computer Technology and
Department of Physics, University of Queensland, Brisbane,
Queensland 4072, Australia} 
\affiliation{Institute for Theoretical
Physics, Kohn Hall, \\
University of California Santa Barbara, Santa
Barbara, CA 93106, USA}

\begin{abstract}
  This paper demonstrates that how well a state performs as an input
  to Grover's search algorithm depends critically upon the
  entanglement present in that state; the more entanglement, the less
  well the algorithm performs.  More precisely, suppose we take a pure
  state input, and prior to running the algorithm apply local unitary
  operations to each qubit in order to maximize the probability
  $P_{\max}$ that the search algorithm succeeds.  We prove that, for
  pure states, $P_{\text{max}}$ is an {\em entanglement monotone}, in
  the sense that $P_{\text{max}}$ can never be decreased by local
  operations and classical communication.
\end{abstract}


\maketitle

\section{Introduction}

%
%

A celebrated result in quantum information
science~\cite{Nielsen00a,Preskill98c} is the discovery of quantum
algorithms able to solve problems faster than any known classical
algorithm.  Three such algorithms are Shor's factoring
algorithm~\cite{Shor94a,Shor97a}, Grover's search
algorithm~\cite{Grover96a,Grover97a}, and algorithms for quantum
simulation (see, for example,~\cite{Nielsen00a}, and references
therein).  However, a satisfactory {\em general} theory of quantum
algorithms is yet to be developed.  Such a theory must address the
question of what makes quantum computers powerful.  No complete answer
to this question has been given, to date, but it is generally believed
that \emph{quantum entanglement} plays a key role.  The purpose of
this paper is to connect the success of Grover's search algorithm with
the amount of entanglement present in the state input to the
algorithm.  

%
%

%
%

In particular, we investigate what physical properties of the initial
state of Grover's algorithm limit the effectiveness of the algorithm.
We show that there is a sense in which {\em the more entanglement is
  present in the initial state, the worse Grover's algorithm
  performs.}  To be more precise, suppose we are given a state
$|\psi\rangle$ and the ability to do local unitary operations on
$|\psi\rangle$ to maximize the probability $P_{\max}(\psi)$ of a
successful run of Grover's algorithm.  The main result of this paper
is to prove that, up to small corrections, $P_{\max}(\psi)$ is an {\em
  entanglement monotone}~\cite{Vidal00b}.  That is, if $|\psi\rangle$
may be transformed into $|\phi\rangle$ by local operations and
classical communication, then we prove that $P_{\max}(\psi) \leq
P_{\max}(\phi)$, again, up to small corrections.  We utilise this
observation to construct an entanglement measure, the \emph{Groverian
  entanglement} of a pure state $\psi$, $G(\psi)$.  We prove that the
Groverian entanglement is, up to small corrections, an entanglement
monotone, and is equivalent to an entanglement measure proposed
previously by Vedral, Plenio, Rippin and Knight~\cite{Vedral97a}.
Thus, this work provides an {\em operational interpretation} for a
multi-party entanglement measure, explicitly connecting that measure
to the success probability of a quantum algorithm.

%
%

The paper is organised as follows.  In Section~\ref{sec:algorithm} we
describe the quantum search algorithm and derive an exact expression
for the maximal success probability for a given initial register
state.  Motivated by this expression, in Section~\ref{sec:groverian}
we introduce the Groverian entanglement, analyse its properties, and
show that it is an entanglement monotone.  Section~\ref{sec:ext}
investigates generalizations of our results to the case of non-qubit
systems, to mixed states, and specializes to the case of bipartite
systems.  Finally, Section~\ref{sec:conc} summarises and discusses our
results, and suggests directions for further research.

\section{The quantum search algorithm}
\label{sec:algorithm}

In this section we review Grover's quantum search algorithm, and
derive an analytic expression for the probability that the algorithm
succeeds when the initial input state is an arbitrary pure state of
$n$ qubits.

Consider a search space $D$ containing $N$ elements.  We assume, for
convenience, that $N = 2^n$, where $n$ is an integer. In this way, we
may represent the elements of $D$ using an $n$-qubit {\em register}
containing their indices, $i=0,\dots,N-1$.  We assume that a subset of
$r$ elements in the search space are marked, that is, they are
solutions to the search problem.  The distinction between the marked
and unmarked elements can be expressed by a suitable function, $f: D
\rightarrow \{0,1\}$, such that $f=1$ for the marked elements, and
$f=0$ for the rest.

Suppose we wish to search the space $D$ to find a marked element.
Phrased in terms of the function $f$, the search for a marked element
becomes a search for an element such that $f=1$.  To solve this
problem on a classical computer one needs to evaluate $f$ for each
element, one by one, until a marked state is found.  Thus, on average,
$\Theta(N)$ evaluations of $f$ are required on a classical computer.
It is one of the most surprising results in quantum information
science that, if we allow the function $f$ to be evaluated
\emph{coherently}, there exists a sequence of unitary operations which
can locate the marked elements using only $O(\sqrt{N/r})$ coherent
queries of $f$ \cite{Grover96a,Grover97a}.  This sequence of unitary
operations is called Grover's quantum search algorithm.

To describe the operation of the quantum search algorithm we first
introduce a register, $\left| x \right\rangle = \left| x_{1} \ldots
  x_{n} \right\rangle$, of $n$ qubits, and an \emph{ancilla} qubit,
$|q\rangle$, to be used in the computation.  It will be convenient to
sometimes used the label ``$q$'' for the ancilla.  We also introduce a
\emph{quantum oracle}, a unitary operator $O$ which functions as a
black box with the ability to \emph{recognize} solutions to the search
problem.  (For more details on how an oracle may be constructed, see
Chapter~6 of~\cite{Nielsen00a}.)  The oracle performs the following
unitary operation on computational basis states of the register,
$\left| x \right\rangle$, and of the ancilla, $\left| q
\right\rangle$:
\begin{equation}\label{eq:bborac}
O \left| x \right\rangle \left| q \right\rangle =
\left| x \right\rangle   \left| q \oplus f(x) \right\rangle
\end{equation}
where $\oplus$ denotes addition modulo 2. This definition may be
uniquely extended, via linearity, to all states of the register
and ancilla.

The oracle recognizes marked states in the sense that if $|x\rangle$
is a marked element of the search space, $f(x) = 1$, the oracle flips
the ancilla qubit from $\left| 0 \right\rangle$ to $\left| 1
\right\rangle$ and vice versa, while for unmarked states the ancilla
is unchanged.  In Grover's algorithm the ancilla qubit is initially
set to the state $(\left| 0 \right> - \left| 1 \right>)/\sqrt{2}$.  It
is easy to verify that, with this choice, the action of the oracle is:
\begin{equation}
O |x\rangle\left( \frac{|0\rangle-|1\rangle}{\sqrt 2} \right)
= (-1)^{f(x)} |x\rangle\left( \frac{|0\rangle-|1\rangle}{\sqrt 2} \right).
\end{equation}
Thus, the only effect of the oracle is to apply a phase of $-1$ if $x$
is a marked state, and no phase change if $x$ is unmarked.  Since the
state of the ancilla does not change, it is conventional to omit it,
and write the action of the oracle as $O|x\rangle =
(-1)^{f(x)}|x\rangle$.  Grover's search algorithm may be summarised as
follows:

\noindent \emph{Algorithm 1:} Grover's quantum search algorithm.

\noindent \emph{Inputs:} (i) a black box oracle $O$, whose action
is defined by Eq.~(\ref{eq:bborac}); (ii) $n+1$ qubits in the
state $|0\rangle^{\otimes n}|0\rangle_q$.

\noindent \emph{Outputs:} a candidate for a marked state, $|s\rangle$.

\noindent \emph{Procedure:}

\begin{enumerate}
\item \label{en:initst} Initialization: Apply a Hadamard gate   
$H = \frac{1}{\sqrt2}\left(\begin{smallmatrix} 1 & 1 \\ 1 & -1
\end{smallmatrix}\right)$ to each qubit in the register, and the gate 
$HX$ to the ancilla, where $X=\left(\begin{smallmatrix} 0 & 1 \\ 1 & 0
  \end{smallmatrix}\right)$ is the {\sc not} gate, 
and we write matrices with respect to the computational basis
($|0\rangle,|1\rangle$).  The resulting state is:
\begin{equation}\label{eq:eta}
\frac{1}{\sqrt{N}}\sum_{x=0}^{2^n-1} |x\rangle
\left(\frac{|0\rangle - |1\rangle}{\sqrt2}\right)_q.
\end{equation}
\label{init:item}
\item Grover Iterations: Repeat the following operation $m$ times,
  where $m$ is an integer whose construction we describe below:
\begin{enumerate}
\item \label{en:rot1} Apply the oracle, which has the effect of
  rotating the marked states by a phase of $\pi$ radians.  Since the
  ancilla is always in the state $(|0\rangle-|1\rangle)/\sqrt 2$ the effect
  of this operation
  may be described by a unitary operator acting only on the register,
  $I_f^{\pi} = \sum_{x} (-1)^{f(x)} | {x} \rangle \langle {x} |$.
\item Rotate all register states by $\pi $ radians around the average
  amplitude of the register state. This is done by (i) applying the
  Hadamard gate to each qubit in the register; (ii) rotating the
  $\left| 00 \ldots 0 \right\rangle$ state 
  of the register by a phase of $\pi$
  radians.  This rotation is similar to 2(a), except for the fact that
  here it is performed on a known state. It takes the form
  $I_{0}^{\pi} = -|0\rangle \langle 0| + \sum_{x\neq 0} |x\rangle
  \langle x|$. (iii) Again applying the Hadamard gate to each qubit
  in the register.
\end{enumerate}
The combined operation on the register
is described by $U_G = H^{\otimes n} I_{0}^{\pi} H^{\otimes n} I_f^\pi$.
\item  Measure the register in the computational basis.
\end{enumerate}
Missing from this description is a value for $m$.  As subsequent
Grover iterations are applied, the amplitudes of the marked states
gradually increase, while the amplitudes of the unmarked states
decrease.  There exists an optimal number, $m$, of iterations at which
the amplitude of the marked states reaches a maximum value, and thus
the probability that the measurement yields a marked state is maximal.
Let us denote this probability by $P$.  It has been
shown~\cite{Grover97a,Boyer98a} that $m$ is bounded above
\begin{equation}\label{eq:optit}
m \le\left\lceil\frac{\pi}{4} \sqrt{\frac{N}{r}}\right\rceil,
\end{equation}
where $r$ is the number of marked states.  The exact value of $m$ as a
function of $N$ and $r$ has been constructed
in~\cite{Boyer98a,Zalka99a}.  Moreover, it has been shown that
Grover's algorithm is optimal in the sense that it is as efficient as
theoretically possible~\cite{Bennett97b}, and that it is possible to
obtain the marked state with very high probability,
$P=1-O(1/\sqrt{N})$, after $m$ iterations~\cite{Boyer98a,Zalka99a}.
Note that $P \approx 1$ only occurs for the specific starting state
described in step~$1$ of Algorithm~$1$, above.  If the Grover
iterations start from an arbitrary state, then $P$ may be bounded away
from $1$~\cite{Biham99a}.

In this paper we are interested in determining what properties of the
initial state of the register are responsible for the efficiency of
the quantum search algorithm.  To this end, we propose modifying the
initialization step, as described by the following hypothetical
situation: Consider $n$ parties (Alice, Bob, Charlie, \dots, Narelle)
sharing a pure quantum state $|\phi\rangle$.  For simplicity, we
initially assume that $|\phi\rangle$ is a state of $n$ qubits, and
each party is in possession of one qubit.  The parties wish to
cooperate in a joint venture in which they use \emph{those particular
  $n$ qubits} to perform a quantum search of the space of $N=2^n$
elements.  The parties are unable to employ any communication
channels.  Prior to the search, each party may perform local unitary
operations on the qubit in their possession.  After they complete the
local processing of their qubits, all parties send (or teleport) their
qubits to the search processing unit.  The only processing available
in this unit is Grover's search iterations and the subsequent
measurement. Thus, the only way the qubits are allowed to interact is
through Grover iterations.

This modified quantum search algorithm, which, with variations, we
study for the remainder of this paper, may be summarised as follows:

\noindent \emph{Algorithm 2:} Modified quantum search.

\noindent \emph{Inputs:} (i) a black box oracle $O$, whose action
is defined by Eq.~(\ref{eq:bborac}); (ii) $n+1$ qubits in the
state $|\phi\rangle|0\rangle_q$.

\noindent \emph{Outputs:} a candidate for a marked state, $|s\rangle$.

\noindent \emph{Procedure:}

\begin{enumerate}
\item \label{en:initst2} Initialization: Apply to the input
  register-ancilla state, $|\phi\rangle|0\rangle_q$, a product of
  arbitrary local operations on the register, $V=U_1 \otimes U_2
  \otimes \cdots \otimes U_n$, and the gate $HX$ on the
  ancilla, where $U_j$ is an arbitrary local unitary gate acting
  on the $j$th qubit.  The resulting state is
\begin{equation}\label{eq:eta2}
|\psi\rangle \otimes \left(\frac{|0\rangle - |1\rangle}{\sqrt2}\right)_q =
V|\phi\rangle\otimes HX|0\rangle_q.
\end{equation}
\label{init:item2}
\item  Grover Iterations: Repeat the following operation $m$ times, where $m$
is chosen as described above:
\begin{enumerate}
\item \label{en:rot2} Rotate the marked states by a phase of $\pi$ radians,
as in Algorithm~$1$.
\item Rotate all register states by $\pi $ radians around the average
amplitude of the register state, as in Algorithm~$1$.
\end{enumerate}
The combined operation on the register
is described by $U_G = H^{\otimes n} I_{0}^{\pi} H^{\otimes n} I_f^\pi$.
\item  Measure the register in the computational basis.
\end{enumerate}

%
%

This modification of Grover's algorithm may appear somewhat {\em ad
  hoc}.  However, as we now explain, the modification allows a
connection between Grover's algorithm and measures of entanglement to
be made.

%
%

The connection follows by asking what is the maximal probability of
success, $P_{\text{max}}$, that a marked element is found, where the
maximization is over all possible local unitary operations in the
initialization step?  We will analyse this question for the case where
there is just a {\em single} marked solution, which we denote $s$, to
the search problem.  We show that in this case $P_{\max}$ is related
to the entanglement present in the initial register state,
$|\phi\rangle$.

To make this assertion more precise, let us write $P_{\max}$ in terms
of the operator $U_G^m$ representing $m$ Grover iterations.  Averaging
uniformly over all $N$ possible values for $s$~\cite{endnote29} we see
that this probability may be written
\begin{equation}\label{eq:Pmaxdef}
P_{\max} = \max_{U_1,\ldots,U_n} \frac{1}{N} \sum_{s=0}^{N-1}
\left| \langle s| U_G^m (U_1 \otimes U_2 \otimes \cdots \otimes
U_n) |\phi\rangle \right|^2,
\end{equation}
where the maximization is over all local unitary operations $U_1,
\ldots, U_n$ on the respective qubits.

To analyse~(\ref{eq:Pmaxdef}) for a general state, $|\phi\rangle$, a
simple trick allows us consider only the action of the Grover
iterations on the equal superposition state $|\eta\rangle = \sum_x
|x\rangle / \sqrt{N}$ which is usually used as the input to Grover's
algorithm.  Applying $m$ Grover iterates to this state yields
\begin{equation}\label{eq:getms}
U_G^m \left| \eta \right> = |s\rangle +
O\left(\frac{1}{\sqrt{N}}\right),
\end{equation}
where the second term is a small correction due to the fact that
Grover's algorithm does not yield a solution with probability $1$, but
rather with probability $1-O(1/\sqrt N)$.  Multiplying this equation
by $(U_G^m)^{\dagger}$ and then taking the Hermitian conjugate gives
\begin{equation}
\langle s| U_G^m
= \langle\eta| + O\left(\frac{1}{\sqrt{N}}\right).
\end{equation}
Substituting into Eq.~(\ref{eq:Pmaxdef}) gives, for a general state
$|\phi\rangle$,
\begin{equation}\label{eq:pmax2}
\begin{split}
P_{\max} = \max_{U_1,\ldots,U_n} \frac{1}{N} \sum_{s=0}^{N-1}
\left| \langle \eta | U_1 \otimes U_2 \otimes \cdots \otimes U_n
|\phi\rangle \right|^2 \\ + O\left(\frac{1}{\sqrt{N}}\right).
\end{split}
\end{equation}
However, $|\eta\rangle$ is a product state, so that $U_1^\dag
\otimes U_2^\dag \otimes \cdots \otimes U_n^\dag|\eta\rangle$ is
another product state. Therefore, the optimization in
Eq.~(\ref{eq:pmax2}) may, equivalently, be expressed as an
optimization over product states,
\begin{equation}\label{eq:pmax3}
P_{\max} = \max_{|e_1, \ldots, e_n\rangle} \left| \langle
e_1,\ldots, e_n|\phi \rangle \right|^2 +
O\left(\frac{1}{\sqrt{N}}\right),
\end{equation}
where the maximization now runs over all product states,
$|e_1,\ldots,e_n\rangle=|e_1\rangle\otimes\cdots|e_n\rangle$, of the
$n$ qubits.  In order for the parties Alice, Bob, Charlie, \dots,
Narelle to achieve this maximum probability when running
Algorithm~$2$, they apply to the joint state $|\phi\rangle$ local
unitary rotations $U_j$ which have the effect of taking $|e_j\rangle$
to $(|0\rangle+|1\rangle)/\sqrt 2$.

This expression, Eq.~(\ref{eq:pmax3}), takes a suggestive form.  Up to
corrections of order $1/\sqrt N$ it depends monotonically on the
maximum of the overlap between all product states and the input state
$|\phi\rangle$~\cite{endnote30}.  If the input state were a product,
$|\phi\rangle =
|u_1\rangle\otimes|u_2\rangle\otimes\cdots\otimes|u_n\rangle$, then
$P_{\text{max}}$ would be equal to one, again, up to small
corrections.  If, alternatively, the input state were not a product
state, it would never be possible for the modified search algorithm to
succeed with probability one.  These observations suggest that
$P_{\text{max}}$ depends, in some way, on the entanglement of the
initial register state, $|\phi\rangle$.

\section{An entanglement measure from the quantum search
algorithm}
\label{sec:groverian}

In the last section we suggested that the maximum success probability,
$P_{\text{max}}$, of Algorithm~$2$, depended on the entanglement of
the initial state of the register.  In this section, we show that
$P_{\max}$ can be used to define an entanglement measure, the {\em
  Groverian entanglement}, for arbitrary pure multiple qubit states.
We show that the Groverian entanglement is closely related to an
entanglement measure introduced previously by Vedral, Plenio, Rippin
and Knight~\cite{Vedral97a} (see also Vedral and
Plenio~\cite{Vedral98a}, and Barnum and Linden~\cite{Barnum01a}).
This connection enables us to understand some properties of the
Groverian entanglement making it a good entanglement measure.

Before defining the Groverian entanglement, we briefly overview some
common approaches taken to the definition of entanglement measures.
Broadly speaking, there are two main approaches, an {\em operational
  approach}, and an {\em axiomatic} approach.  In the operational
approach~\cite{Bennett96a}, measures of entanglement are related to
physical tasks that one can perform with a quantum state, like quantum
communication.  The axiomatic approach (see, for
example,~\cite{Vedral97a,Vidal00b}) starts from desirable axioms that
a ``good'' entanglement measure should satisfy, and then attempts to
construct such measures.

%
%

The Groverian entanglement is an example of an entanglement measure
defined in operational terms, namely, how well a state serves as the
input to Algorithm~$2$.  We define the Groverian entanglement of a
state $|\psi\rangle$ by:
\begin{equation}\label{eq:entgrov}
G (\psi) \equiv \sqrt{1 - P_{\max}}.
\end{equation}
Note that we will freely interchange the notations $|\psi\rangle$ and
$\psi$.  Since $P_{\max}$ takes values in the range $0 \le P_{\max}
\le 1$, it follows that $0 \le G (\psi) \le 1$.  However, it is not
immediately clear that $G(\psi)$ is a good measure of entanglement.
We show that this is the case by using the results of the previous
section to connect $G(\psi)$ to a measure of entanglement introduced
in~\cite{Vedral97a}, following the axiomatic approach.

%
%

To demonstrate the connection between the Groverian entanglement
and~\cite{Vedral97a}, we substitute Eq.~(\ref{eq:pmax3}) into
Eq.~(\ref{eq:entgrov}), and move the maximization outside the square
root, where it becomes a minimization.  Neglecting terms of $O(1/\sqrt
N)$ this gives
\begin{equation}\label{eq:intstep1}
G(\psi) = \min_{|e_1,\ldots,e_n\rangle} \sqrt{1- 
F^2(e_1 \otimes \cdots \otimes e_n,\psi)},
\end{equation}
where $F(\cdot,\cdot)$ is the {\em
  fidelity}~\cite{Uhlmann76a,Jozsa94c,Nielsen00a}, defined in general
by $F(\rho,\sigma) \equiv \mbox{tr} \sqrt{\sqrt \rho \sigma \sqrt
  \rho}$.  Special cases of interest are the pure state fidelity,
$F(a,b) = |\langle a|b\rangle|$, and the case where one state is pure
and one state is mixed, $F(\sigma,a) = \langle
a|\sigma|a\rangle^{1/2}$.  We now show that we can extend the range of
the minimization in Eq.~(\ref{eq:intstep1}) to a minimization over the
space ${\cal S}$ of \emph{all} separable density matrices, that is,
density matrices which can be written in the form $\sigma = \sum_j p_j
\rho^1_j \otimes \ldots \otimes \rho^n_j$,
\begin{equation} \label{eq:formula}
G(\psi) = \min_{\sigma \in {\cal S}} \sqrt{1- F^2(\sigma,\psi)}.
\end{equation}
To see this, simply note that by linearity of $F^2(\sigma,\psi)$ in
$\sigma$, and convexity of ${\cal S}$, the maximal value of
$F^2(\sigma,\psi)$, and thus the minimum in
$\sqrt{1-F^2(\sigma,\psi)}$, can always be obtained at an extreme
point of ${\cal S}$, that is, when $\sigma$ is a pure product state.

%
%
The expression Eq.~(\ref{eq:formula}) for the Groverian entanglement
should be compared with the following definition of an entanglement
measure, introduced in~\cite{Vedral97a} by Vedral, Plenio, Rippin and
Knight~\cite{endnote31}:
\begin{eqnarray}
E(\psi) \equiv 2-2\max_{\sigma \in {\cal S}} F(\sigma,\psi).
\end{eqnarray}
This definition is essentially equivalent to ours, in that $G(\psi)$
is a monotonic function of $E(\psi)$, and vice versa.  Vedral {\em et
  al} introduced their definition motivated primarily by axiomatic
concerns; we have shown that, in fact, there is a close connection
between this measure and the utility of the state as an input to
Grover's algorithm.

%
%
We now briefly describe several useful properties of the Groverian
entanglement.  The proofs are the same as those given
in~\cite{Vedral97a} (see also~\cite{Vedral98a,Barnum01a}); what is new
is the connection between this measure of entanglement and Grover's
algorithm.  It is clear that $G(\psi)=0$ iff $|\psi\rangle$ is a
product state, and that local unitary operations on the qubits leave
$G(\psi)$ invariant.  What is more surprising in the context of
Grover's algorithm, and is the main result of this paper, is that
$G(\psi)$ is an {\em entanglement monotone}.  That is, $G(\psi)$
cannot be increased by local operations and classical communication:

%
%

{\bf Theorem:} Let $|\psi\rangle$ and $|\phi\rangle$ be $n$-qubit pure
states such that it is possible to transform $|\psi\rangle$ to
$|\phi\rangle$ by local operations on the qubits, and classical
communication.  Then $G(\psi) \geq G(\phi)$, up to corrections of
order $1/\sqrt{N}$.

%
%

This theorem has the remarkable implication that the probability
$P_{\max}$ of success for our modified Grover's algorithm can {\em
  never decrease} under local operations and classical communication.
The proof of the theorem follows easily by rewriting
Eq.~(\ref{eq:formula}) in terms of the {\em Bures
  metric}~\cite{Bures69a}, which is defined by~\cite{endnote32}:
\begin{equation}
B(\rho,\sigma) \equiv \sqrt{1-F^2(\rho,\sigma)},
\end{equation}
which results in
\begin{equation} \label{eq:grover_bures}
G(\psi) = \min_{\sigma \in {\cal S}} B(\sigma,\psi).
\end{equation}
Suppose $|\psi\rangle$ can be transformed into $|\phi\rangle$ by a
process of local operations and classical communication, whose effect
is represented by the quantum operation~\cite{Nielsen00a} ${\cal E}$.
Let $\sigma$ be the state for which the minimum in
Eq.~(\ref{eq:grover_bures}) is achieved, $G(\psi) =
B(\sigma,\psi)$.  It can be shown~\cite{Barnum96a} that the
Bures distance between two states can never be increased by a quantum
operation, so
\begin{eqnarray}
G(\psi) & = & B(\sigma,\psi) \\
 & \geq & B({\cal E}(\sigma),{\cal E}(|\psi\rangle \langle \psi|)) \\
 & = & B({\cal E}(\sigma),\phi).
\end{eqnarray}
But $\sigma$ is separable, so ${\cal E}(\sigma)$ is also
separable, since it can be obtained by local operations and classical
communication from $\sigma$.  Thus
\begin{eqnarray}
G(\psi) \geq B({\cal E}(\sigma),\phi) \geq G(\phi),
\end{eqnarray}
which completes the proof that $G(\cdot)$ is an entanglement monotone.


\section{Extensions of the Groverian entanglement}
\label{sec:ext}

%
%
In this section we investigate three scenarios generalizing the
earlier results about $n$-qubit pure state entanglement.
Subsection~\ref{subsec:dimensionality} addresses systems whose
subsystems are not qubits but instead have arbitrary (finite)
dimensionality.  Subsection~\ref{subsec:bipartite} specializes to the
case of a {\em bipartite} quantum system, where the two subsystems
have arbitrary finite dimensionalities.  Finally, in
Subsection~\ref{subsec:mixed} we consider whether the Groverian
entanglement is a good measure of entanglement for mixed states.

\subsection{Groverian entanglement for subsystems of arbitrary dimensionality}
\label{subsec:dimensionality}

%
%
As described earlier, Algorithm~$2$ is applied to a system of $n$
qubits, and thus the Groverian entanglement is only defined for such a
system.  However, with a small modification the algorithm we described
can be extended to the case of $n$ systems of arbitrary finite
dimensionalities, $d_1,d_2,\ldots,d_n$.

%
%
The only change is in the inversion about the average, step 2(b).  To
achieve the analogous operation, we need to find a replacement for the
Hadamard gate.  Suppose $V_j$ is any $d_j \times d_j$ unitary operator
such that $V_j|0\rangle = \sum_{k=0}^{d_j-1} |k\rangle / \sqrt{d_j}$,
where $|0\rangle,\ldots,|d_j-1\rangle$ forms an orthonormal basis for
the state of the $j$th system.  For example, $V_j$ could be the matrix
representation of the Fourier transform over the integers modulo
$d_j$.  Then the inversion about the average can be achieved by (i)
applying the operation $V_j$ to each system; (ii) rotating the $\left|
  00 \ldots 0 \right\rangle$ state of the register by a phase of $\pi$
radians.  This rotation takes the form $I_{0}^{\pi} = -|0\rangle
\langle 0| + \sum_{x\neq 0} |x\rangle \langle x|$; (iii) applying the
inverse operation $V_j^{\dagger}$ to each system.

%
%
With this modification, the Grover iterate can be used to perform
quantum searches using systems of arbitrary dimensionality.
Proceeding as before, we find that Eq.~(\ref{eq:pmax3}) holds even for
systems of arbitrary dimensionality, that is,
\begin{equation} \label{eq:pmax4}
P_{\max} = \max_{|e_1, \ldots, e_n\rangle} \left| \langle
e_1,\ldots, e_n|\phi \rangle \right|^2 +
O\left(\frac{1}{\sqrt{N}}\right),
\end{equation}
Similarly, if we define the Groverian entanglement by $G(\psi) \equiv
\sqrt{1 - P_{\max}}$ then the same argument as before shows that the
Groverian entanglement is an entanglement monotone, up to corrections
of $O(1/\sqrt N)$, and can thus be regarded as a good measure of
entanglement for composite systems of arbitary dimensionality.

\subsection{Two-party Groverian entanglement} 
\label{subsec:bipartite}

In this section we specialize our study of the Groverian entanglement
to bipartite quantum systems and derive an analytic expression for the
Groverian entanglement in that case.  We suppose that the two
component systems have arbitrary finite dimensionalities, $d_1$ and
$d_2$.  In the bipartite case the optimization in Eq.~(\ref{eq:pmax4})
is equivalent to the maximization of the fidelity,
\begin{equation}\label{eq:fidl1}
F(U\otimes V|0\rangle_A |0\rangle_B,\phi)
\end{equation}
where we use the fact that any product state may be written as a
product of two local unitaries operating on some fiducial state
$|0\rangle_A|0\rangle_B$.  This problem has been considered
in~\cite{Barnum99a,Vidal00a}, where it was shown that the solution may
be obtained in terms of the \emph{Schmidt
  decomposition}~\cite{Nielsen00a} of $|\phi\rangle$,
\begin{equation}\label{eq:schmidt}
|\phi\rangle = \sum_i \sqrt{p_i}|u^i\rangle_A|v^i\rangle_B,
\end{equation}
where $|u^i\rangle$ and $|v^i\rangle$ are each orthonormal sets of
vectors, and the Schmidt coefficients $\sqrt{p_i}$ are non-negative
real numbers.  \cite{Barnum99a,Vidal00a} showed that the maximum
occurs when $U\otimes V|0\rangle_A|0\rangle_B =
|u^i\rangle_A|v^i\rangle_B$ where $i$ is chosen so that $\sqrt{p_i} =
\sqrt{p_{\max}}$ is the maximal Schmidt coefficient.  Substituting
into Eq.~(\ref{eq:pmax4}) gives
\begin{equation}
G(\psi) = \sqrt{1-p_{\max}}.
\end{equation}

Thus, for a bipartite system the Groverian entanglement is equivalent
to a well-known entanglement monotone~\cite{Nielsen99a,Vidal00b}, the
square of the largest Schmidt coefficient.  Indeed, for the case of
two qubits, $G(\psi)$ is equivalent to the usual asymptotic measure of
pure state entanglement~\cite{Bennett96c,Bennett96a,Popescu97a}, the
von Neumann entropy of the reduced density operator for either qubit,
$S=-\mbox{tr}(\rho_A\log\rho_A)$.  The relationship between the two
quantities is $S = h(G^2(\psi))$, where $h(x) =
-x\log_2x-(1-x)\log_2(1-x)$ is the binary entropy.

\subsection{The Groverian entanglement for mixed states}
\label{subsec:mixed}

%
%
We have defined the Groverian entanglement and investigated its
properties for the special case of pure state inputs to Grover's
algorithm.  How does the analogous measure behave for mixed states?
Is it still a good measure of entanglement?  In this subsection we
briefly consider these questions.  We show that the natural
generalization to mixed states is not a good measure of entanglement,
and discuss other possible ways of generalizing the Groverian
entanglement to mixed states.

%
%
Suppose a mixed state $\rho$ is used as the input in Algorithm~$2$,
replacing the pure state $|\phi\rangle$.  Then it is not difficult to
show that the corresponding maximal probablity of success is given by
\begin{equation}  \label{eq:pmax5}
P_{\max} = \max_{|e_1, \ldots, e_n\rangle} \langle
e_1,\ldots, e_n|\rho|e_1,\ldots, e_n\rangle +
O\left(\frac{1}{\sqrt{N}}\right),
\end{equation}
which is the linear extension of the expression in
Eq.~(\ref{eq:pmax3}) to a general density matrix.  Suppose we define
\begin{eqnarray}
G(\rho) \equiv \sqrt{1-P_{\max}}.
\end{eqnarray}
For pure states this agrees with the earlier definition of the
Groverian entanglement.

%
%
Suppose $\rho = \rho_1 \otimes \ldots \otimes \rho_n$, and that
$\lambda_j$ is the largest eigenvalue of $\rho_j$.  Then from
Eq.~(\ref{eq:pmax5}), $P_{\max} = \lambda_1\cdot \lambda_2 \cdot
\ldots \cdot \lambda_n$, and thus
\begin{eqnarray}
G(\rho_1 \otimes \ldots \otimes \rho_n) = \sqrt{1-\prod_{j=1}^n \lambda_j}.
\end{eqnarray}
In the case when $\rho_1,\ldots,\rho_n$ are pure states, all the
$\lambda_j = 1$, and $G(\rho) = 0$.  However, when the $\rho_j$ are
mixed, the values of $G(\rho)$ may span the entire range from
$G(\rho)$'s minimal value of $0$, right up to its maximal possible
value of $\sqrt{1-1/N}$.  It follows that $G(\rho)$ cannot be an
entanglement monotone.

%
%
From these observations we conclude that $G(\rho)$ is not a good
measure of entanglement for mixed states.  The essential problem is
that $G(\rho)$ is linear in $\rho$, and many states that we ordinarily
think of as not being entangled can be represented as a mixture of
entangled states.  For example, the completely mixed state $I \otimes
I/4$ of two qubits can be written as an equal mixture of maximally
entangled states.  By linearity, $G(I\otimes I/4)$ therefore takes the
same value as for a maximally entangled state.

%
%

Is there any sensible way of resolving this difficulty with mixed
states?  At present, we are not aware of any natural resolution that
preserves the elegant operational interpretation of the Groverian
entanglement.  It is interesting to note, however, that Vedral {\em et
  al}'s~\cite{Vedral97a} proposed measure of entanglement applied
equally well to either pure or mixed states.  In particular, for a
general mixed state $\rho$ of a composite system one can define
\begin{eqnarray}
\tilde G(\rho) \equiv \min_{\sigma \in {\cal S}} \sqrt{1-F^2(\rho,\sigma)},
\end{eqnarray}
where the minimization is over all {\em separable} states $\sigma$ of
the system, and $F(\rho,\sigma)$ is the fidelity, as defined earlier.
This is a generalization of our measure for pure states, however we
have not succeeded in obtaining a good operational interpretation of
$\tilde G(\rho)$ along lines similar to the pure state case.  Another
possible resolution, following a similar line of thought
to~\cite{Barnum01a}, is to define
\begin{eqnarray}
\hat G(\rho) \equiv \min \sum_j p_j G(\psi_j),
\end{eqnarray}
where the minimum is over all ensembles $\{ p_j, |\psi_j\rangle \}$
such that $\rho = \sum_j p_j |\psi_j\rangle \langle \psi_j|$.  It is
not difficult to show that $\hat G(\rho)$ is an entanglement monotone,
locally unitarily invariant, and is equal to zero if and only if
$\rho$ is separable.  However, once again, a good operational
interpretation of $\hat G(\rho)$ is presently unknown to us.

\section{Summary, discussion, and future directions}
\label{sec:conc}

%
%

In this paper we have investigated the relationship between the
success probability of a modified form of Grover's quantum search
algorithm, and the amount of entanglement present in the initial state
used for the algorithm.  We have proposed an entanglement measure for
$n$-party pure states, the Groverian entanglement, based on the
maximal success probability of the algorithm.  Furthermore, we showed
that the Groverian entanglement is essentially equivalent to a measure
of entanglement introduced by Vedral, Plenio, Rippin and
Knight~\cite{Vedral97a}, and used this to argue that the Groverian
entanglement and $P_{\max}$ are entanglement monotones.

%
%

Our work suggests several directions for future research.  It would be
interesting to investigate other variants of Grover's algorithm,
including:
\begin{enumerate}
\item Allowing multiple solutions in the search space, rather than a
  single solution, as we have considered.  
\item Replacing the two Hadamard transforms in the Grover iterate by
  an arbitrary unitary transform $U$ and its inverse $U^{\dagger}$,
  respectively.
\item Tracking the evolution of the entanglement present in
  intermediate stages of the algorithm.  Investigations along these
  lines, but in a somewhat different context, have been reported
  in~\cite{Vedral01a,Meyer01a,Latorre01a}.
\item Determining the effect noise has on the performance of the
  algorithm, and entanglement measures derived from the algorithm.
\end{enumerate}
It would also be interesting to investigate other quantum algorithms,
such as Shor's algorithm, quantum simulation, and adiabatic quantum
computation~\cite{Farhi01a}.  We hope that by pursuing such
investigations insight will be obtained into the fundamental question
of what makes quantum computers powerful, and further elucidate the
role entanglement plays in quantum information processing.

\section{Acknowledgements}

Thanks to Nick Bonesteel for discussions that stimulated this work,
and to Guifre Vidal for providing comments on the manuscript.
Especial thanks to Jennifer Dodd for an exceptionally thorough reading
of the manuscript, and many helpful suggestions on how to improve it.
OB thanks the Centre for Quantum Computer Technology and the
Department of Physics at the University of Queensland for hospitality
during a sabbatical leave when this work was done.  This research was
supported, in part, by an Australian Postgraduate Award to TJO, and by
the National Science Foundation under Grant No.~PHY99-07949.
Preliminary work at the Hebrew University in Jerusalem was supported
by the EU fifth framework program Grant No.~IST-1999-11234.


\begin{thebibliography}{28}
\bibitem{endnote30}{The correction term in Eq.\protect \nobreakspace
    {}(\ref {eq:getms}) may actually be removed by an additional
    modification of Grover's algorithm in which the rotations in the
    Grover iterate are not by $\pi $ radians, but rather by some other
    angle close to $\pi $. For the sake of clarity we do not concern
    ourselves with this issue, and will mostly be content to ignore
    the corrections of order $1/\protect \sqrt N$ in any case.}
\bibitem{endnote31}{Note that the conventions for the fidelity used
    in\protect \nobreakspace {}\cite {Vedral97a} differ slightly from
    ours. The quantity they call the fidelity is the square of the
    quantity we call fidelity. We have rewritten their expressions to
    take this difference into account.}
\bibitem{endnote32}{The Bures metric is sometimes written with an
    extra constant factor $\protect \sqrt 2$ out the front. Note
    that\protect \nobreakspace {}\cite {Vedral97a,Vedral98a}
    erroneously describes the function $2-2F(\rho ,\sigma )$ as the
    Bures metric. If we choose $\psi = |0\delimiter "526930B , \phi =
    (|0\delimiter "526930B +|1\delimiter "526930B )/\protect \sqrt
    {2}, \tau = |1\delimiter "526930B $ then the triangle inequality
    $2-2F(\psi ,\tau ) \leq 2-2F(\psi ,\phi )+ 2-2F(\phi ,\tau )$ is
    violated, so this function cannot be a metric.}
\end{thebibliography}

\end{document}